\documentclass[aps,pre,twocolumn]{revtex4}

\usepackage{graphicx}
\usepackage{bm}
\usepackage{bbm}
\usepackage{amsmath}
\DeclareMathOperator{\Tr}{Tr}

\begin{document}

\title{Theory of Liquid Crystal Elastomers:  From Polymer Physics to Differential Geometry}
\author{Thanh-Son Nguyen}
\author{Jonathan V. Selinger}

\date{December 14, 2016}

\begin{abstract}
In liquid crystal elastomers, the orientational order of liquid crystals is coupled with elastic distortions of crosslinked polymer networks.  Previous theoretical research has described these materials through two different approaches:  a neoclassical theory based on the liquid crystal director and the deformation tensor, and a geometric elasticity theory based on the difference between the actual metric tensor and a reference metric.  Here, we connect those two approaches using a formalism based on differential geometry.  Through this connection, we determine how both the director and the geometry respond to a change of temperature.
\end{abstract}

\maketitle

\section{Introduction}

Liquid crystal elastomers are composed of liquid-crystalline mesogenic units covalently bonded to crosslinked polymer networks~\cite{Warner2003}.  Because of this structure, they combine the orientational order of liquid crystals with the elasticity of polymer networks.  Any distortion of the polymer network affects the orientational order, and conversely, any change in the magnitude or direction of the orientational order affects the shape of the polymer network.  For that reason, liquid crystal elastomers are being studied extensively for applications as actuators or shape-changing materials.

Early theoretical work on liquid crystal elastomers was based on polymer physics.  In a series of influential papers, summarized in the textbook~\cite{Warner2003}, Warner and Terentjev generalized the classical theory of rubber elasticity into the ``neoclassical'' theory of liquid crystal elastomers.  In this generalization, they describe the elastomer as a set of cross-links connected by polymer chains, which are modeled as \emph{anisotropic} random walks.  The free energy of the material then becomes the entropic free energy of the polymer chains, and it can be expressed as the trace of combination of tensors representing the anisotropic step lengths at the time of cross-linking, the anisotropic step lengths at the current time, and the elastic distortion of the polymer network.

More recent theoretical research on liquid crystal elastomers has been based on geometric elasticity theory~\cite{Modes2010,Modes2011a,Modes2011b,Modes2012,Cirak2014,Aharoni2014,Pismen2014,Mostajeran2015,Modes2015,Modes2016,Mostajeran2016,Plucinsky2016}.  In this mathematical approach, the basic concept is that any material has an actual metric tensor $g$, which represents the distances between nearby material points, and a reference metric tensor $\bar{g}$, which represents the most favorable distances between such points.  The difference between these tensors defines the strain on the material, and hence gives the elastic energy.  This mathematical approach is well suited to liquid crystal elastomers, because many experiments change the natural shape of the material by changing some conditions (such as temperature or optical illumination), and observe how the material responds.  The change in natural shape can be expressed by a change in the reference metric tensor $\bar{g}$.  Indeed, liquid crystal elastomers provide an ideal way to tune the reference metric tensor, because they allow $\bar{g}$ to be anisotropic.

One important difference between these two theoretical approaches is how they treat the liquid crystal director field.  In the first approach, the director field is a distinct thermodynamic variable, which is only weakly coupled to the underlying polymer network; it is able to realign in complex ways in response to elastic distortions.  This approach is appropriate for lightly crosslinked elastomers, with only small interactions between orientational and elastic degrees of freedom.  By contrast, in the second approach, the director field is fixed with respect to the polymer network at the time of sample preparation, through a procedure called ``blueprinting''~\cite{Modes2011b}.  After the sample is prepared, the director cannot change; only the magnitude of nematic order can change.  This approach is appropriate for strongly crosslinked elastomers, sometimes called liquid crystal glasses, which have strong interactions between orientational and elastic degrees of freedom.

The purpose of this paper is to connect those two approaches.  We begin with the neoclassical theory, based on the Warner-Terentjev trace formula, supplemented by their semisoft elastic term.  We re-express this theory in the language of differential geometry, developing a method that can describe the geometry and director at the time of crosslinking, as well as the geometry and director at the current time.  We then minimize the free energy over the current director, to obtain the optimal orientation as a function of current geometry, initial geometry, and initial director.  For the case of a strongly crosslinked elastomer, the current director is locked to its optimal orientation.  In that case, the neoclassical theory reduces to the geometric elasticity theory, with a reference metric $\bar{g}$ expressed in terms of the initial geometry and initial director.

By connecting these approaches, we obtain several insights.  First, we can see that the theory of liquid-crystal elastomers is analogous to recent research on swellable gel sheets, which develop complex 3D shapes in response to programmable swelling patterns~\cite{Klein2007,Kamien2007,Efrati2009,Sharon2010,Dias2011}.  Second, we can model the distinction between liquid crystal elastomers and liquid crystal glasses, depending on how strongly the director is locked into an orientation determined by the elastic distortion.  Third, we can analyze the set of degenerate configurations in ideal elastomers with soft elasticity, and see how the degeneracy is broken by the non-ideal semi-soft elasticity.

The plan of this paper is as follows.  In Sec.~II, we develop a geometric formalism to describe the geometry and director at the time of crosslinking and at the current time.  In Sec.~III, we construct the free energy by translating the Warner-Terentjev neoclassical theory into the geometric formalism.  In Sec.~IV, we consider the difference between weakly and strongly crosslinked elastomers, and calculate the director orientation in the limit of strong crosslinking (or liquid crystal glasses).  Finally, in Sec.~V, we determine the favored metric tensor $\bar{g}$ as a function of temperature in a strongly crosslinked elastomer.

\section{Geometric Formalism}

A key concept in the Warner-Terentjev theory of liquid crystal elastomers is that we must keep track of two states of the same material:  the state at the time of crosslinking and the state at the current time.  In this section, we set up a geometric formalism to describe both of these states in terms of the \emph{same} internal coordinate system.  This internal coordinate system is not necessarily Cartesian; it might be 2D polar, 3D cylindrical or spherical, or anything else.

When we set up this formalism, we consider three possible dimensionalities for the problem.

\emph{Case 1:} 2D material embedded in 2D plane.

\begin{figure}
\includegraphics[width=2.5in]{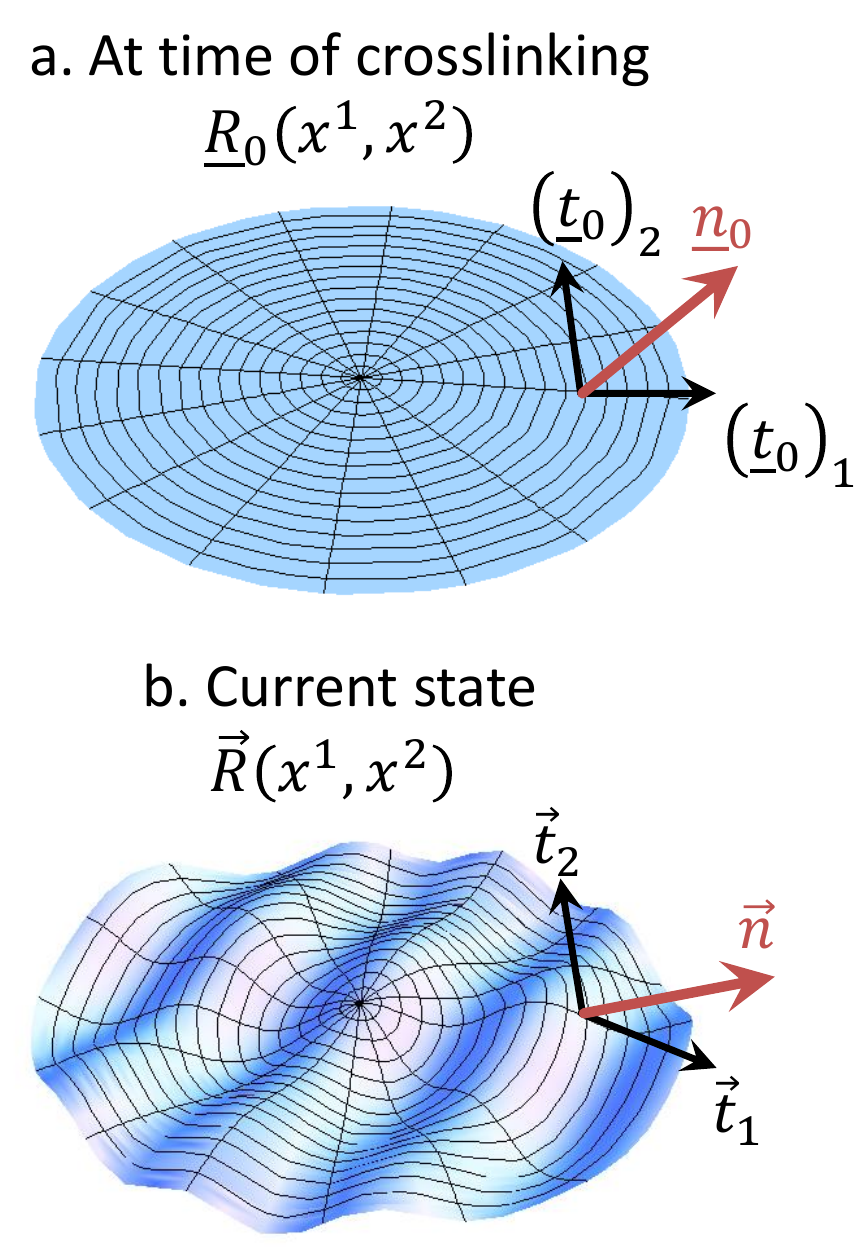}
\caption{Schematic illustration of the material at the time of crosslinking, compared with the material at the current time.  This picture represents case 2, which is a 2D material embedded in 3D space, described using polar coordinates.}
\label{schematicgeometry}
\end{figure}

\emph{Case 2:} 2D material embedded in 3D space.  (This is the case illustrated in Fig.~\ref{schematicgeometry}.)

\emph{Case 3:} 3D material embedded in 3D space.

\subsection{Geometry at time of crosslinking}

At the time of crosslinking, the system is described by the position $\underline{R}_0$ of any material point as a function of the internal coordinates $x^\alpha$.  For cases 1 and 2, this position is a 2D vector $\underline{R}_0 (x^1,x^2)$, and the index $\alpha$ ranges over 1 and 2.  For case 3, this position is a 3D vector $\underline{R}_0 (x^1,x^2,x^3)$, and the index $\alpha$ ranges over 1, 2, and 3.

Any vector or tensor field in the material must be represented in terms of local basis vectors.  The conventional choice of covariant basis vectors is $(\underline{t}_0)_\alpha=\partial_\alpha\underline{R}_0$.  The covariant metric tensor is then $(g_0)_{\alpha\beta}=(\underline{t}_0)_\alpha \cdot (\underline{t}_0)_\beta$, and its inverse is the contravariant metric tensor $(g_0)^{\alpha\beta}$.  The contravariant basis vectors are $(\underline{t}_0)^\alpha=(g_0)^{\alpha\beta}(\underline{t}_0)_\beta$.

The material may have nematic orientational order at the time of crosslinking.  The magnitude of nematic order can be characterized by $s_0=r_0 -1$, where $r_0$ is the Warner-Terentjev anisotropy parameter.  Note that $s_0=0$ is isotropic and $s_0 >0$ is nematic.  The nematic director is represented by the unit vector $\underline{n}_0$.  The director is not necessary uniform; it might vary with position in a complex way designed by the experimenter.  For cases 1 and 2 it is a 2D vector $\underline{n}_0 (x^1,x^2)$; for case 3 it is a 3D vector $\underline{n}_0 (x^1,x^2,x^3)$.

The polymer step length tensor can be written as
\begin{equation}
\underline{\underline{\ell}}_0=a_0\left[\underline{\underline{I}}+s_0 \underline{n}_0 \underline{n}_0\right].
\end{equation}
We will refer to this type of expression, which is independent of coordinate system, as an abstract tensor.  The step length tensor has contravariant components in the selected coordinate system
\begin{equation}
(\ell_0)^{\alpha\beta}=(\underline{t}_0)^\alpha \cdot a_0\left[\underline{\underline{I}}+s_0 \underline{n}_0 \underline{n}_0\right] \cdot (\underline{t}_0)^\beta .
\end{equation}
We will refer to this type of expression, where an abstract tensor is placed between two basis vectors to obtain the tensor components, as a sandwich parameterization.

\subsection{Geometry at current time}

At the current time, the system is described by the position $\vec{R}$ of any material point as a function of the same internal coordinates $x^\alpha$ defined above.  For case 1 this position is a 2D vector $\vec{R}(x^1,x^2)$, for case 2 it is a 3D vector $\vec{R}(x^1,x^2)$, and for case 3 it is a 3D vector $\vec{R}(x^1,x^2,x^3)$.

To represent any vector or tensor field in the current geometry, the conventional choice of covariant basis vectors is $\vec{t}_\alpha=\partial_\alpha\vec{R}$.  The covariant metric tensor is then $g_{\alpha\beta}=\vec{t}_\alpha \cdot \vec{t}_\beta$.

The current magnitude of nematic order can be characterized by $s$, and the current director is represented by $\vec{n}$.  For case 1 the director is a 2D unit vector $\vec{n}(x^1,x^2)$, for case 2 it is a 3D unit vector $\vec{n}(x^1,x^2)$, and for case 3 it is a 3D unit vector $\vec{n}(x^1,x^2,x^3)$.

The current polymer step length tensor can be written as the abstract tensor
\begin{equation}
\stackrel{\leftrightarrow}{\ell}=a\left[\stackrel{\leftrightarrow}{I}+s\vec{n}\vec{n}\right],
\end{equation}
and its inverse is
\begin{equation}
\stackrel{\leftrightarrow}{\ell}^{-1}=a^{-1}\left[\stackrel{\leftrightarrow}{I}-\frac{s}{1+s}\vec{n}\vec{n}\right].
\end{equation}
In the selected coordinate system, the covariant components of the inverse step length tensor are
\begin{equation}
(\ell^{-1})_{\alpha\beta}=\vec{t}_\alpha \cdot a^{-1}\left[\stackrel{\leftrightarrow}{I}-\frac{s}{1+s}\vec{n}\vec{n}\right] \cdot \vec{t}_\beta
\end{equation}
as a sandwich parameterization.

Note that we have invented a typographical convention where underlines denote vectors and tensors in the system at the time of crosslinking, and arrows above the symbols denote vectors and tensors in the system at the current time.  This convention serves as a typographical reminder of which space is described.

\section{Free Energy}

The physical problem can now be stated as follows.  Suppose we know the state of the material at the time of crosslinking.  In particular, we know the shape given by $\underline{R}_0 (x^\alpha)$, and hence the tangent vectors $(\underline{t}_0)_\alpha$ and metric $(g_0)_{\alpha\beta}$, as well as the nematic order parameter $s_0$ and the director field $\underline{n}_0 (x^\alpha)$.  We now change temperature or illuminate the material, so that the nematic order parameter changes to a new value $s\not=s_0$.  (We might also apply stresses that couple to the shape, or electric or magnetic fields that couple to the director.)  We would like to predict the new shape $\vec{R}(x^\alpha)$ and the new director field $\vec{n}(x^\alpha)$.  We emphasize that this is a physical problem, not a mathematical problem; there is no mathematical transformation that maps $\underline{R}_0$ or $\underline{n}_0$ onto $\vec{R}$ or $\vec{n}$.  Rather, we must calculate them by minimizing the free energy.

For an ideal, homogeneous elastomer at fixed volume, Warner and Terentjev~\cite{Warner2003} argue that the free energy is given by the trace formula for soft elasticity,
\begin{equation}
F_\text{soft}=\frac{1}{2}\mu\Tr[\bm{\lambda}^T \bm{\ell}^{-1} \bm{\lambda} \bm{\ell}_0],
\end{equation}
where $\mu$ is the shear modulus and $\bm{\lambda}=\partial\vec{R}/\partial\underline{R}_0$ is the deformation tensor.  For a non-ideal, heterogeneous elastomer, the free energy has an additional term for semisoft elasticity
\begin{equation}
F_\text{semi}=\frac{1}{2}\mu\alpha\Tr[\bm{\lambda}^T \bm{n}\bm{n} \bm{\lambda} (\bm{I}-\bm{n}_0 \bm{n}_0)],
\end{equation}
where $\alpha>0$ is the semisoft parameter, which is proportional to $s_0 s$ for small $s_0$ and $s$, because the semisoft term is only defined if there is nematic order at the time of crosslinking and at the current time.  For an elastomer that can change volume, these terms must be supplemented by a contribution
\begin{equation}
F_\text{bulk}=\frac{1}{2}B[\det(\bm{\lambda})-1]^2,
\end{equation}
where $B$ is the bulk modulus.

Each of these terms can be transformed into the language of differential geometry, using the notation defined in the previous section.  The soft elastic term becomes
\begin{eqnarray}
F_\text{soft}&=&\frac{1}{2}\mu(\ell^{-1})_{\alpha\beta}(\ell_0)^{\beta\alpha}\nonumber\\
&=&\frac{\mu a_0}{2a}\left[\vec{t}_\alpha \cdot \left[\stackrel{\leftrightarrow}{I}-\frac{s}{1+s}\vec{n}\vec{n}\right]
\cdot \vec{t}_\beta\right]\times\\
&&\qquad\times\left[(\underline{t}_0)^\beta \cdot \left[\underline{\underline{I}}+s_0 \underline{n}_0 \underline{n}_0\right] \cdot (\underline{t}_0)^\alpha\right].
\nonumber
\end{eqnarray}
Note that the expression $\vec{t}_\beta(\underline{t}_0)^\beta$ corresponds to the deformation tensor $\bm{\lambda}$.  The semisoft term becomes
\begin{eqnarray}
F_\text{semi}&=&\frac{1}{2}\mu\alpha(\vec{n}\vec{n})_{\alpha\beta}(\underline{\underline{I}}-\underline{n}_0 \underline{n}_0)^{\beta\alpha}\\
&=&\frac{1}{2}\mu\alpha\left[\vec{t}_\alpha \cdot (\vec{n}\vec{n}) \cdot \vec{t}_\beta\right]
\left[(\underline{t}_0)^\beta \cdot (\underline{\underline{I}}-\underline{n}_0 \underline{n}_0) \cdot (\underline{t}_0)^\alpha\right].
\nonumber
\end{eqnarray}
Finally, the bulk modulus term becomes
\begin{eqnarray}
F_\text{bulk}&=&\frac{1}{8}B\left[\frac{\det(g_{\alpha\beta})-\det((g_0)_{\alpha\beta})}{\det((g_0)_{\alpha\beta})}\right]^2\\
&=&\frac{1}{8}B\left[\frac{\det(\vec{t}_\alpha \cdot\vec{t}_\beta)-\det((\underline{t}_0)_\alpha \cdot(\underline{t}_0)_\beta)}{\det((\underline{t}_0)_\alpha \cdot(\underline{t}_0)_\beta)}\right]^2.\nonumber
\end{eqnarray}
The total free energy is therefore defined as
\begin{equation}
F_\text{total}=F_\text{soft}+F_\text{semi}+F_\text{bulk}.
\end{equation}

We emphasize that the total free energy is a function of the current $\vec{t}_\alpha$ vectors, which describe the shape of the material, and the current $\vec{n}$ vector, which represents the liquid-crystal director.  It is not expressed in terms of the deviations of these vectors from any reference state.  Of course, the free energy includes information about the state of the material at the time of crosslinking, through the $(\underline{t}_0)_\alpha$ and $\underline{n}_0$ vectors.  However, these vectors are defined in a different space, possibly even in a space with different dimension.  The vectors in different spaces can never be subtracted from each other.  They are only compared through their expressions in the coordinate system, which is common to both spaces.  Because the free energy involves invariant expressions, it does not depend on the specific choice of coordinate system.

\section{Director Orientation in Strongly Crosslinked Elastomers}

Now that we have the free energy, we can suggest a way to think about the difference between weakly and strongly crosslinked liquid crystal elastomers.  In a weakly crosslinked elastomer, the shape and director are two separate degrees of freedom, which are both represented in $F_\text{total}(\vec{t}_\alpha,\vec{n})$.  Of course these degrees of freedom are coupled in the free energy, but the coupling is small.  The director $\vec{n}$ might not always be oriented in a way that minimizes the free energy for the current shape $\vec{t}_\alpha$, either because of external perturbations or because of thermal fluctuations.

By contrast, in a strongly crosslinked elastomer (i.~e. liquid crystal glass), the shape and director are not separate degrees of freedom.  Rather, the director is locked to the current shape of the material.  As a result, $\vec{n}$ is always at the minimum of $F_\text{total}(\vec{t}_\alpha,\vec{n})$ for the current $\vec{t}_\alpha$.

One possible counter-argument to this way of thinking is:  The Warner-Terentjev trace formula is derived using certain assumptions of polymer physics, particularly with the assumption that the polymer strands are long Gaussian chains between the crosslinks.  In a strongly crosslinked elastomer, this assumption might no longer be valid.  Hence, one cannot rely on the trace formula for the free energy.

Our response to this counter-argument is:  Suppose we just want to derive the free energy for a strongly crosslinked elastomer as a function of $\vec{t}_\alpha$ and $\vec{n}$ based on symmetry, without making any assumptions about the microscopic physics.  This free energy would have exactly the same couplings of $\vec{t}_\alpha$ and $\vec{n}$ with $(\underline{t}_0)_\alpha$ and $\underline{n}_0$ as in $F_\text{soft}$ and $F_\text{semi}$, because these are the lowest-order couplings that are allowed by symmetry.  The only difference is that we would not refer to $\stackrel{\leftrightarrow}{\ell}$ and $\underline{\underline{\ell}}_0$ as step length tensors; we would just say that these are some arbitrary tensors allowed by symmetry.  For that reason, we will continue to use the same mathematics, and just modify the words as needed.

For the rest of this article, we will consider a strongly crosslinked elastomer.  Based on the argument above, the director $\vec{n}$ must be at the minimum of $F_\text{total}(\vec{t}_\alpha,\vec{n})$ for the current $\vec{t}_\alpha$.  Hence, we should calculate this minimum explicitly to determine the director orientation.

The director-dependent terms in $F_\text{total}$ can be written as 
\begin{equation}
F_\text{total}=\text{const}-\frac{\mu a_0 s(1-\alpha')}{2a(1+s)}\vec{n}\cdot\stackrel{\leftrightarrow}{\mathcal{M}}\cdot\vec{n},
\label{ftotalfunctionofn}
\end{equation}
where $\stackrel{\leftrightarrow}{\mathcal{M}}$ is the tensor defined by
\begin{equation}
\stackrel{\leftrightarrow}{\mathcal{M}}=\vec{t}_\beta (\underline{t}_0)^\beta \cdot \left[\underline{\underline{I}}
+\left(\frac{s_0+\alpha'}{1-\alpha'}\right) \underline{n}_0 \underline{n}_0\right] \cdot (\underline{t}_0)^\alpha \vec{t}_\alpha.
\label{Mtensor}
\end{equation}
For convenience, we are using a rescaled semisoft parameter $\alpha'$ defined by
\begin{equation}
\alpha=\frac{a_0}{a}\frac{s}{1+s}\alpha' .
\end{equation}
Because $\alpha$ is proportional to $s_0 s$, we must have $\alpha'$ proportional to $s_0$ for small $s_0$.

From Eq.~(\ref{ftotalfunctionofn}), the minimum of $F_\text{total}$ occurs when $\vec{n}$ is the eigenvector of $\stackrel{\leftrightarrow}{\mathcal{M}}$ corresponding to the largest eigenvalue.  To determine the eigenvector, we must look at the form of $\stackrel{\leftrightarrow}{\mathcal{M}}$ in Eq.~(\ref{Mtensor}).  Note that $\stackrel{\leftrightarrow}{\mathcal{M}}$ has two terms, and \emph{neither} of these terms is isotropic.  The two terms favor alignment of $\vec{n}$ in different directions, and hence compete with each other.  Let us consider each of these terms separately.

The first term in $\stackrel{\leftrightarrow}{\mathcal{M}}$ is
\begin{equation}
\stackrel{\leftrightarrow}{\mathcal{M}}_1 = \vec{t}_\beta (\underline{t}_0)^\beta \cdot (\underline{t}_0)^\alpha \vec{t}_\alpha
=\vec{t}_\beta (g_0)^{\beta\alpha} \vec{t}_\alpha.
\end{equation}
In terms of the deformation tensor $\bm{\lambda}$, it reduces to
\begin{equation}
\stackrel{\leftrightarrow}{\mathcal{M}}_1 = \bm{\lambda} \bm{\lambda}^T
= \stackrel{\leftrightarrow}{I}+2 \stackrel{\leftrightarrow}{\epsilon}_\text{Almansi},
\end{equation}
where $\stackrel{\leftrightarrow}{\epsilon}_\text{Almansi}$ is the Almansi strain tensor, i.e.\ the strain tensor defined in the current configuration space.  Hence, $\stackrel{\leftrightarrow}{\mathcal{M}}_1$ tends to align $\vec{n}$ along the main strain axis.  If the material is crosslinked in the isotropic phase (with $s_0 =0$ and $\alpha'=0$), then this is the only contribution to $\stackrel{\leftrightarrow}{\mathcal{M}}$, and hence $\vec{n}$ will exactly align along the main strain axis.

The second term in $\stackrel{\leftrightarrow}{\mathcal{M}}$ is
\begin{equation}
\stackrel{\leftrightarrow}{\mathcal{M}}_2=\left(\frac{s_0 +\alpha'}{1-\alpha'}\right)\left[\vec{t}_\beta (\underline{t}_0)^\beta \cdot \underline{n}_0 \right]
\left[\underline{n}_0 \cdot (\underline{t}_0)^\alpha \vec{t}_\alpha \right].
\end{equation}
In terms of the deformation tensor $\bm{\lambda}$, it reduces to
\begin{equation}
\stackrel{\leftrightarrow}{\mathcal{M}}_2=\left(\frac{s_0 +\alpha'}{1-\alpha'}\right)\left[\bm{\lambda}\underline{n}_0 \right]
\left[\bm{\lambda}\underline{n}_0 \right].
\end{equation}
Hence, $\stackrel{\leftrightarrow}{\mathcal{M}}_2$ tends to align $\vec{n}$ along $\bm{\lambda}\underline{n}_0$, which is a transformed version of the director at the time of crosslinking.  If this were the \emph{only} term in $\stackrel{\leftrightarrow}{\mathcal{M}}$, then $\vec{n}$ would be a unit vector in that direction, $\vec{n}=(\bm{\lambda}\underline{n}_0)/|\bm{\lambda}\underline{n}_0|$.

Note that the alignment of $\vec{n}$ favored by $\stackrel{\leftrightarrow}{\mathcal{M}}_2$ is exactly the alignment suggested by Ref.~\cite{Cirak2014}.  Hence, we agree with this expression for the alignment in two situations.  The first situation is if $\stackrel{\leftrightarrow}{\mathcal{M}}_2$ dominates over $\stackrel{\leftrightarrow}{\mathcal{M}}_1$, which occurs if $s_0 \gg1$, i.e.\ the nematic order at the time of crosslinking is very large.  The second situation is if the alignment favored by the strain $\stackrel{\leftrightarrow}{\epsilon}_\text{Almansi}$ happens to be the same as the alignment favored by $\bm{\lambda}\underline{n}_0$.  In general, however, the favored alignment of $\vec{n}$ must be a compromise between those two directions.

\subsection*{Example}

\begin{figure}
\includegraphics[width=3.375in]{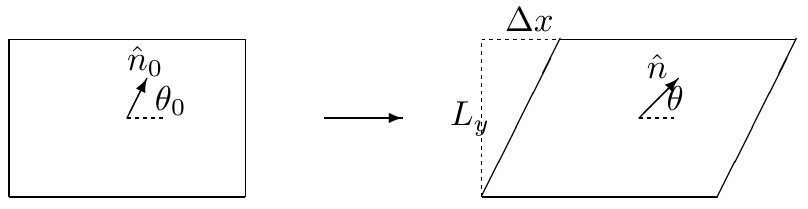}
\caption{Shear deformation of a rectangular elastomer sample.}
\label{exampledeformation}
\end{figure}

As a specific example, consider the director realignment induced by a simple shear deformation of a rectangular elastomer sample, as shown in Fig.~\ref{exampledeformation}.  Suppose that the initial director $\underline{n}_0$ is aligned at an angle of $\theta_0$ with respect to the laboratory $x$-axis.  The material is then subjected to a small shear strain of $\epsilon=\Delta x/L_y$, while the magnitude of nematic order remains unchanged $s=s_0$.  In response to the shear, the new director $\vec{n}$ aligns at an angle $\theta$ with respect to the laboratory $x$-axis.  By working out the $\stackrel{\leftrightarrow}{\mathcal{M}}$ tensor explicitly, and then diagonalizing it, we find
\begin{equation}
\theta=\theta_0 - \frac{\epsilon}{2}\left[1-\left(1+\frac{2(1-\alpha')}{s_0 +\alpha'}\right)\cos2\theta_0\right]
+\mathcal{O}(\epsilon^2).
\label{epsilonseries}
\end{equation}

This example is the same as the example discussed in Ref.~\cite{Nguyen2015}, using a continuum elastic formalism based on the the relative rotation coupling terms $D_1$ and $D_2$~\cite{Warner2003}.  The result is exactly the same, provided that we identify the ratio
\begin{equation}
\frac{D_2}{D_1}=1+\frac{2(1-\alpha')}{s_0 +\alpha'}.
\end{equation}

To interpret this result, we can consider two limits.  First, if the nematic order is very strong ($s_0 \to\infty$), then $D_2 =D_1$.  In this limit, $\stackrel{\leftrightarrow}{\mathcal{M}}_2$ dominates over $\stackrel{\leftrightarrow}{\mathcal{M}}_1$, and hence the director alignment is given by $\vec{n}=(\bm{\lambda}\underline{n}_0)/|\bm{\lambda}\underline{n}_0|$.  Second, if if the nematic order is very weak ($s_0 \to 0$ and $\alpha'\to 0$), then $D_2 \gg D_1$.  Here, $\stackrel{\leftrightarrow}{\mathcal{M}}_1$ dominates over $\stackrel{\leftrightarrow}{\mathcal{M}}_2$, and hence the director aligns along the main strain axis ($+45^\circ$ for any $\epsilon>0$,  $-45^\circ$ for any $\epsilon<0$), regardless of $\theta_0$.  In this limit, the power series in $\epsilon$ in Eq.~(\ref{epsilonseries}) breaks down, but explicit diagonalization of $\stackrel{\leftrightarrow}{\mathcal{M}}$ gives $\theta=\pm45^\circ$.

\section{Calculation of Reference Metric}

We now continue the calculation for strongly crosslinked elastomers by assuming that $\vec{n}$ goes to the orientation that minimizes $F_\text{total}$.  As shown above, this is the eigenvector corresponding to the largest eigenvalue of $\stackrel{\leftrightarrow}{\mathcal{M}}$.  Hence, the product $\vec{n}\cdot\stackrel{\leftrightarrow}{\mathcal{M}}\cdot\vec{n}$ is equal to the largest eigenvalue of $\stackrel{\leftrightarrow}{\mathcal{M}}$, which we will call $e_\text{max}$.  In terms of this eigenvalue, the free energy becomes
\begin{eqnarray}
F_\text{total}&=&\frac{\mu a_0}{2a} g_{\alpha\beta}
\left[(\underline{t}_0)^\beta \cdot
\left[\underline{\underline{I}}+s_0 \underline{n}_0 \underline{n}_0\right] \cdot (\underline{t}_0)^\alpha\right]\nonumber\\
&&-\frac{\mu a_0 s(1-\alpha')}{2a(1+s)}e_\text{max}\\
&&+\frac{1}{8}B\left[\frac{\det(g_{\alpha\beta})-\det((g_0)_{\alpha\beta})}{\det((g_0)_{\alpha\beta})}\right]^2.\nonumber
\end{eqnarray}

For case 1 (2D material in 2D plane) or case 2 (2D material in 3D space), we can calculate the largest eigenvalue explicitly as
\begin{equation}
e_\text{max}=\frac{1}{2}\left[\Tr{\stackrel{\leftrightarrow}{\mathcal{M}}}+\sqrt{2\Tr\left(\stackrel{\leftrightarrow}{\mathcal{M}}^2\right)-\left(\Tr\stackrel{\leftrightarrow}{\mathcal{M}}\right)^2}\right],
\label{emax}
\end{equation}
where
\begin{align}
\Tr{\stackrel{\leftrightarrow}{\mathcal{M}}}=
&g_{\alpha\beta} (\underline{t}_0)^\beta \cdot \left[\underline{\underline{I}}+\left(\frac{s_0 +\alpha'}{1-\alpha'}\right)\underline{n}_0 \underline{n}_0\right] \cdot (\underline{t}_0)^\alpha ,\\
\Tr\left(\stackrel{\leftrightarrow}{\mathcal{M}}^2\right)=&g_{\alpha\beta} (\underline{t}_0)^\beta \cdot \left[\underline{\underline{I}}
+\left(\frac{s_0 +\alpha'}{1-\alpha'}\right)\underline{n}_0 \underline{n}_0\right] \cdot (\underline{t}_0)^\gamma\times\nonumber\\
&\times g_{\gamma\delta}
(\underline{t}_0)^\delta \cdot \left[\underline{\underline{I}}+\left(\frac{s_0 +\alpha'}{1-\alpha'}\right)\underline{n}_0 \underline{n}_0\right] \cdot (\underline{t}_0)^\alpha .
\end{align}
For case 3 (3D material in 3D space), the expression for $e_\text{max}$ is similar but more complicated.  Hence, for the rest of this section, we will consider only cases 1 and 2.

Note that this expression for the free energy depends on the current shape of the material \emph{only} through the metric $g_{\alpha\beta}$, not through the individual vectors $\vec{t}_\alpha$.  This is an important feature of the problem, which is physically reasonable.  After we minimize over $\vec{n}$, the free energy must depend only on the distances between points in the material, given by the metric, not on the orientation of the material in the embedding space, given by $\vec{t}_\alpha$.

We now want to minimize the free energy over the metric tensor $g_{\alpha\beta}$.  For that calculation, it is convenient to represent $g_{\alpha\beta}$ as a sandwich parameterization,
\begin{equation}
g_{\alpha\beta}={\underline{t}_0}_\alpha \cdot c \left(\underline{\underline{I}}+d\underline{m}\underline{m}\right)\cdot{\underline{t}_0}_\beta.
\label{gsandwich}
\end{equation}
where $c$ is an overall scale factor, $d$ indicates the magnitude of strain along some axis, and $\underline{m}$ is a unit vector that indicates the main axis of the strain.  Note that $\underline{m}$ is defined in the plane at the time of crosslinking, and is at an angle $\phi$ from $\underline{n}_0$, so that $\underline{m}\cdot\underline{n}_0 =\cos\phi$.  This parameterization of $g_{\alpha\beta}$ has three degrees of freedom ($c$, $d$, and $\phi$), which is appropriate for a symmetric $2\times2$ tensor.

With this parameterization for $g_{\alpha\beta}$, the free energy becomes
\begin{eqnarray}
F_\text{total}(c,d,\phi)&=&\frac{\mu a_0}{2a}c\left[2+d+s_0 (1+d\cos^2 \phi) \right] \nonumber\\
&&-\frac{\mu a_0 s(1-\alpha')}{2a(1+s)}e_\text{max}\\
&&+\frac{1}{8}B\left[c^2 (1+d)-1\right]^2 ,\nonumber
\end{eqnarray}
where $e_\text{max}$ is given by Eq.~(\ref{emax}) and
\begin{align}
\Tr{\stackrel{\leftrightarrow}{\mathcal{M}}}=
& c\left[(2+d)+\left(\frac{s_0 +\alpha'}{1-\alpha'}\right)(1+d\cos^2 \phi)\right],\\
\Tr\left(\stackrel{\leftrightarrow}{\mathcal{M}}^2\right)=& c^2 \biggl[(2+2d+d^2)\nonumber\\
&+2\left(\frac{s_0 +\alpha'}{1-\alpha'}\right)\left[1+d(2+d)\cos^2 \phi\right]\nonumber\\
&+\left(\frac{s_0 +\alpha'}{1-\alpha'}\right)^2 (1+d\cos^2 \phi)^2
\biggr].
\end{align}

\begin{figure*}
(a) Soft and semisoft modes when the sample is cooled from the time of crosslinking\\
\includegraphics[height=2.76in]{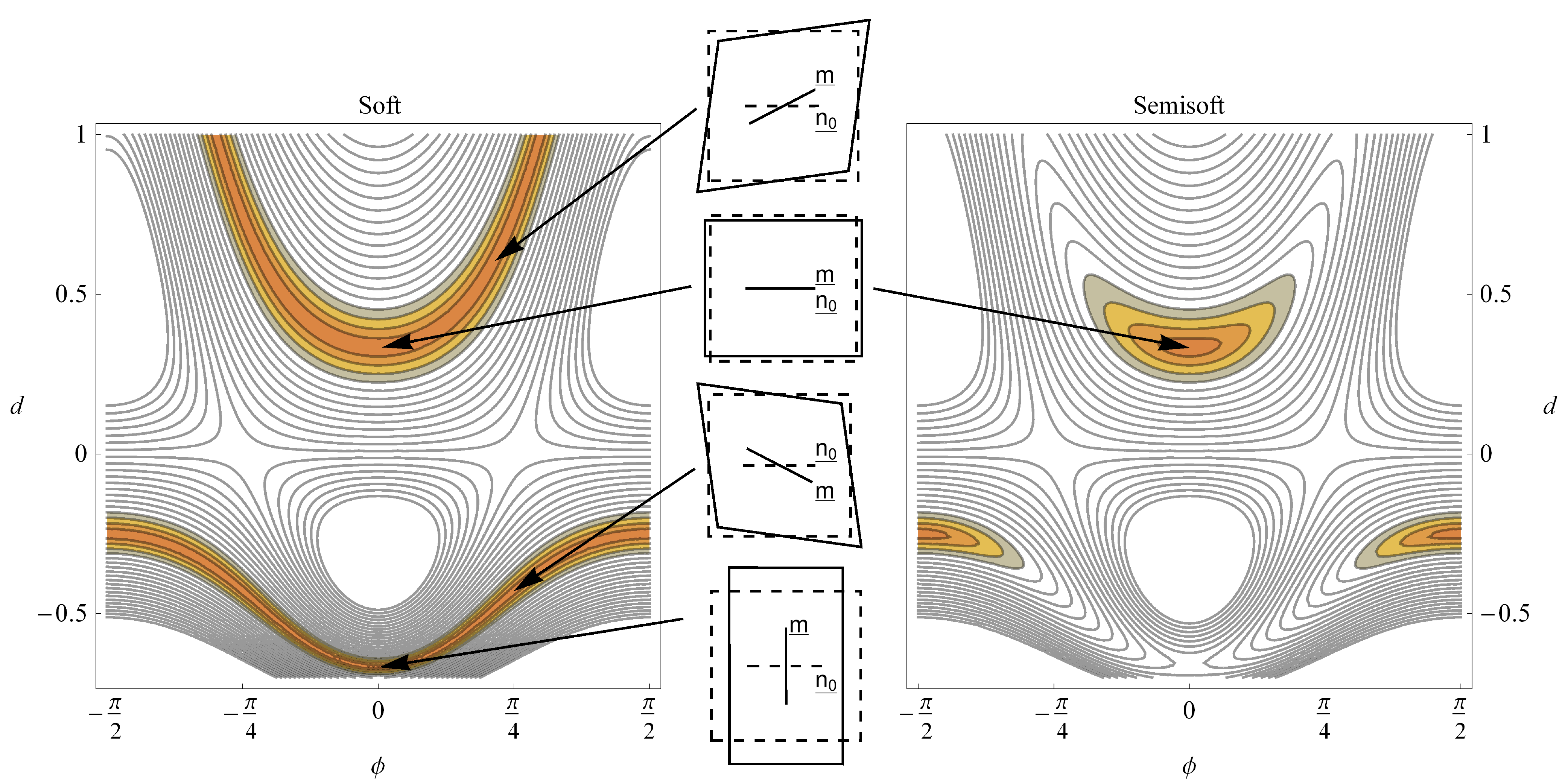}\\
(b) Soft and semisoft modes when the temperature is unchanged from the time of crosslinking\\
\includegraphics[height=2.76in]{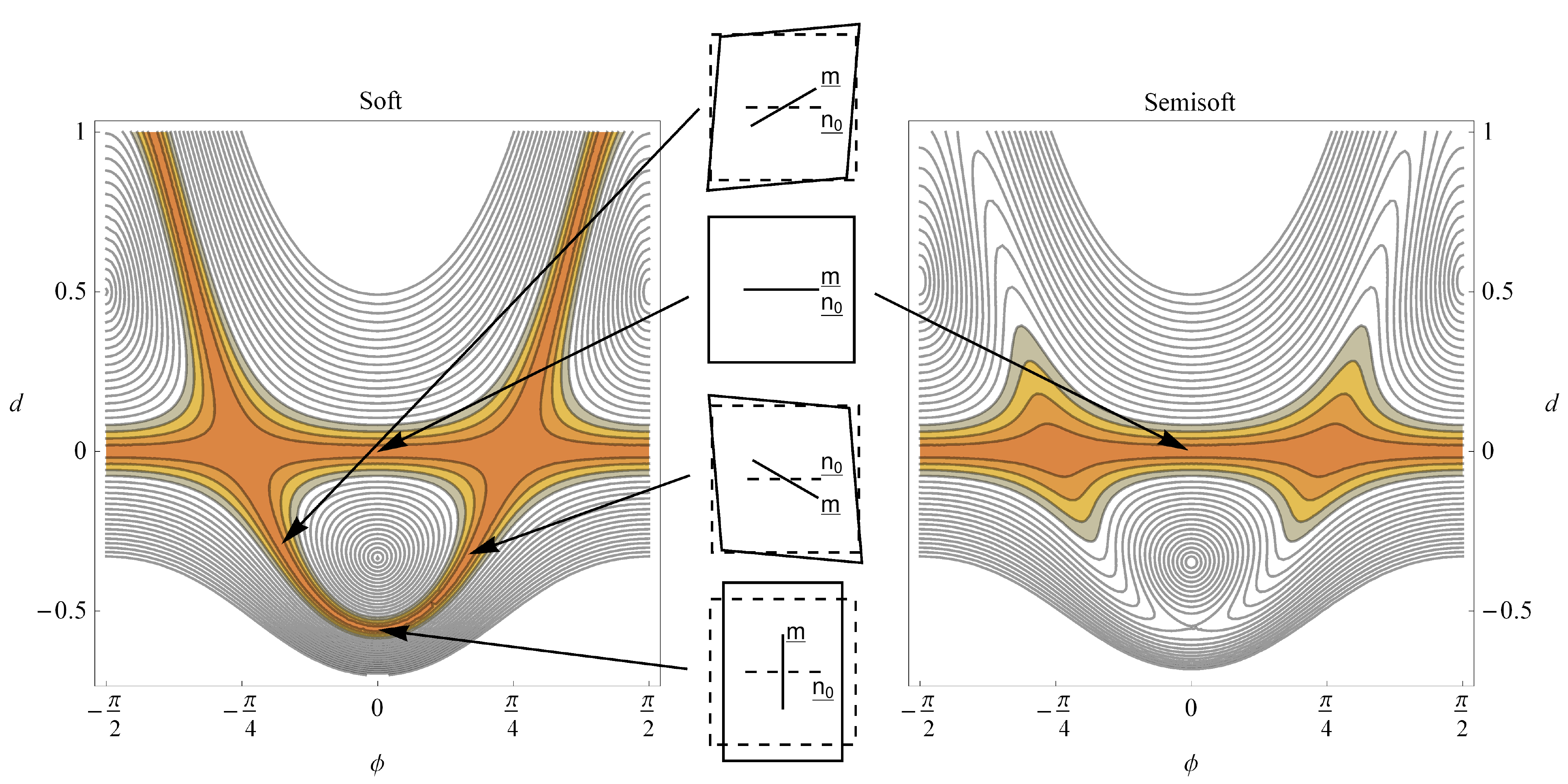}\\
(c) Soft and semisoft modes when the sample is heated from the time of crosslinking\\
\includegraphics[height=2.76in]{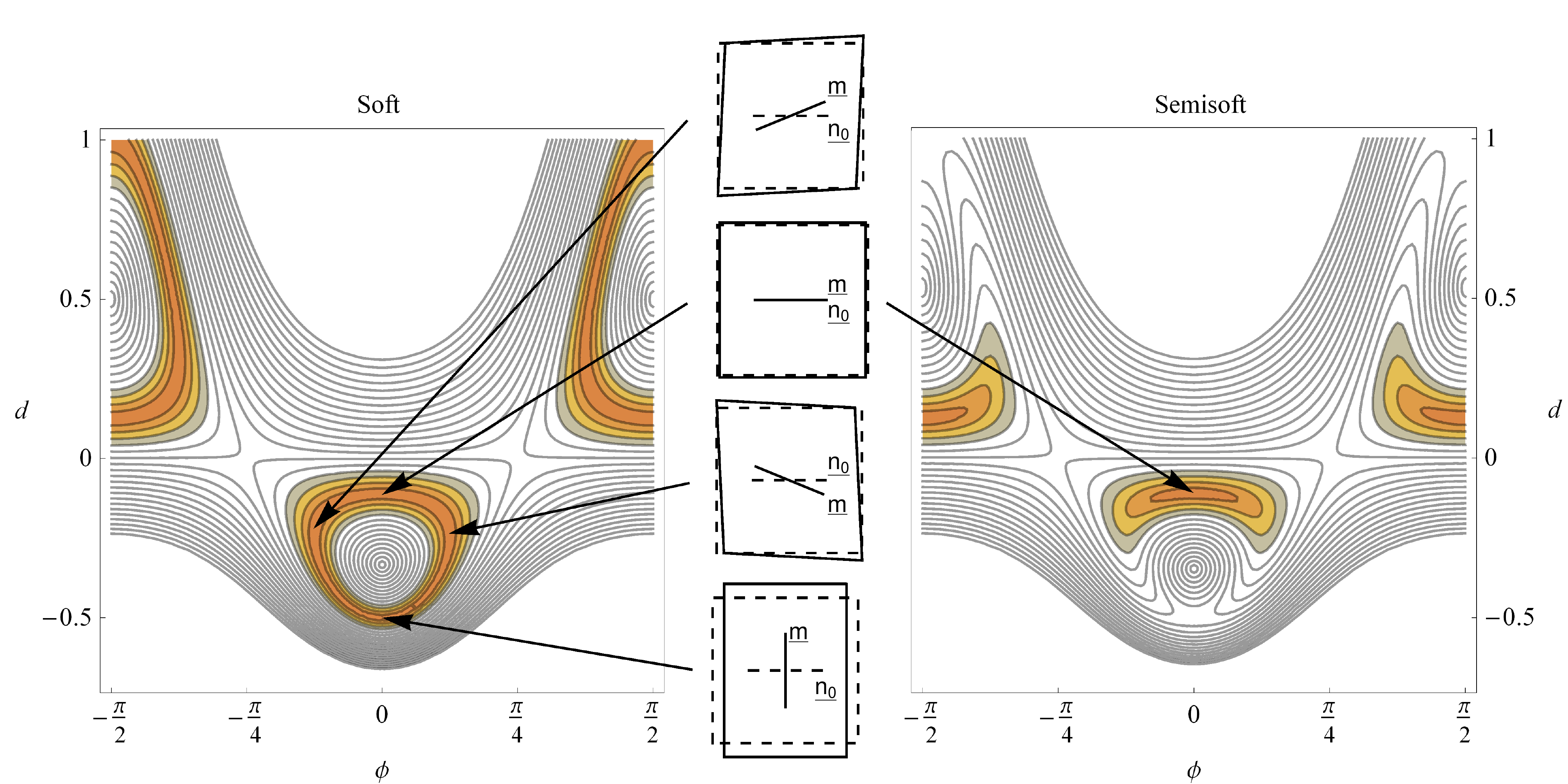}
\caption{The soft (on the left with $\alpha' = 0$) and semisoft (on the right with $\alpha'$ = 0.02) energy contours as functions of the coefficient $d$ and the angle $\phi$ between $\hat n_0$ and $\hat m$.)}
\label{contour}
\end{figure*}

To understand the implications of this free energy function, suppose the material is incompressible, so that $B\to\infty$ and hence $c=(1+d)^{-1/2}$.  In that limit, the free energy reduces to a function of two variables $F_\text{inc}(d,\phi)$.  Furthermore, suppose the material is ideal and homogeneous, so that the semisoft parameter $\alpha'=0$.  The free energy then becomes
\begin{align}
&F^\text{soft}_\text{inc}(d,\phi)=\\
&\frac{\mu a_0}{4a(1+s)(1+d)^{1/2}} \biggl[(2+s)\left[2+d+s_0 (1+d\cos^2 \phi) \right] \nonumber\\
&-s\sqrt{d^2+2s_0 d[-1+(2+d)\cos^2 \phi]+s_0^2 (1+d\cos^2 \phi)^2}\biggr]\nonumber
\end{align}
This soft free energy density admits not just a single minimum, but rather a degenerate set of deformations at the same minimum energy.  By contrast, as the semisoft terms come into play, $\alpha'\not=0$, the degeneracy is broken and a single minimum is chosen as the lowest energy state.  The specific form of the soft modes depends on the temperature difference between the current state and the state at the time of crosslinking. 

As a first example, in Fig.~\ref{contour}(a), suppose the sample is cooled and the order parameter increases from its value at the time of crosslinking, $s > s_0$.  For a soft sample with $\alpha' =0$, there are two continuous sets of minima, as shown by the darkest contours in the left side of the figure.  These two contours represent the same set of deformations, because expanding along $\hat m$ is the same as contracting perpendicular to $\hat m$. Because of the energy degeneracy, there is no energy cost when the sample configuration changes continuously along either dark contour. By contrast, when the semisoft term is introduced, $\alpha' \neq 0$, it singles out one particular configuration as the unique minimum, as shown in the right side of the figure.  In this configuration, the sample stretches along the initial director, with
\begin{equation}
\phi = 0, \qquad d = \frac{s - s_0}{1 + s_0},
\end{equation}
or equivalently it contracts perpendicular to the initial director.

As a second example, in Fig.~\ref{contour}(b), suppose there is no temperature change from the time of crosslinking, so that $s = s_0$.  Clearly the state with no deformation, $d=0$, is a trivial solution for the reference metric, in both the soft and semisoft cases.  In the soft case, we \emph{also} have an additional set of minima at
\begin{equation}
d = - \frac{s_0 (2 + s_0) \cos 2 \phi}{ \left( 1 + s_0 \cos \phi \right)^2},
\end{equation}
along which the free energy is the same as no deformation at all. This is the same phenomenon of soft elasticity as depicted in Ref.~\cite{Warner2003}, Fig.~7.3.

Finally, in Fig.~\ref{contour}(c), suppose the sample is heated from the time of crosslinking, so that $s < s_0$.  For a soft sample with $\alpha' =0$, we can still observe continuous sets of degenerate minima, as shown by the darkest contours on the left side.  As in part~(a), these two contours actually represent the same deformations.  Unlike part~(a), not all values of $\phi$ are available for soft deformation.  When the semisoft term is introduced, $\alpha' \neq 0$, the continuous set of degenerate minima reduces to a unique minimum, as shown on the right side of the figure. 

We can now put together our results for an incompressible, semisoft elastomer in case 1 (2D material in 2D plane) or case 2 (2D material in 3D space).  The lowest free energy occurs when the metric is
\begin{align}
\label{gfinal}
&\bar{g}_{\alpha\beta}={\underline{t}_0}_\alpha \cdot \sqrt{\frac{1+s_0}{1+s}} \left[\underline{\underline{I}}
+\left(\frac{s-s_0}{1+s_0}\right)\underline{n}_0\underline{n}_0\right]\cdot{\underline{t}_0}_\beta\\
&={\underline{t}_0}_\alpha \cdot \left[\sqrt{\frac{1+s}{1+s_0}}\underline{n}_0\underline{n}_0
+\sqrt{\frac{1+s_0}{1+s}}\underline{n}_0^\perp \underline{n}_0^\perp \right]\cdot{\underline{t}_0}_\beta,\nonumber\\
&={\underline{t}_0}_\alpha \cdot \left[\sqrt{\frac{r}{r_0}}\underline{n}_0\underline{n}_0
+\sqrt{\frac{r_0}{r}}\underline{n}_0^\perp \underline{n}_0^\perp \right]\cdot{\underline{t}_0}_\beta\nonumber\\
&=\sqrt{\frac{r}{r_0}}(\underline{n}_0\cdot{\underline{t}_0}_\alpha)(\underline{n}_0\cdot{\underline{t}_0}_\beta)
+\sqrt{\frac{r_0}{r}}(\underline{n}_0^\perp\cdot{\underline{t}_0}_\alpha)(\underline{n}_0^\perp\cdot{\underline{t}_0}_\beta)\nonumber
\end{align}
where $r=1+s$ is the anisotropy parameter of Warner and Terentjev.  Thus, in the most favorable state, the material extends along $\underline{n}_0$ by a factor of $(r/r_0)^{1/2}$, contracts along $\underline{n}_0^\perp$ by the same factor, and this shape change is transformed into the selected coordinate system to form a metric tensor.  This physical result is consistent with the neoclassical theory of liquid crystal elastomers, as in Ref.~\cite{Warner2003}.  The calculated ground state plays the role of the reference metric in the geometric formulation of elasticity theory, as in Refs.~\cite{Klein2007,Efrati2009}, and hence we call it $\bar{g}_{\alpha\beta}$.

If the material reaches a state with the metric $\bar{g}_{\alpha\beta}$, we can then calculate the favored director $\vec{n}$.  We showed earlier that the favored director is the eigenvector of $\stackrel{\leftrightarrow}{\mathcal{M}} = \stackrel{\leftrightarrow}{\mathcal{M}}_1 + \stackrel{\leftrightarrow}{\mathcal{M}}_2$ from Eq.~(\ref{Mtensor}), with the eigenvalue $e_\text{max}$ from Eq.~(\ref{emax}).  By explicit construction, we can see that
\begin{equation}
\vec{n}=\frac{\vec{t}_\alpha (\underline{t}_0)^\alpha \cdot \underline{n}_0}{|\vec{t}_\beta (\underline{t}_0)^\beta \cdot \underline{n}_0|}
=\frac{\bm{\lambda}\underline{n}_0}{|\bm{\lambda}\underline{n}_0|}
\label{nfinal}
\end{equation}
is a normalized eigenvector with the correct eigenvalue, and hence it is the favored director.  In other words, when the material has the metric $\bar{g}_{\alpha\beta}$, there is no conflict between $\stackrel{\leftrightarrow}{\mathcal{M}}_1$ and $\stackrel{\leftrightarrow}{\mathcal{M}}_2$; both tensors have the same eigenvector given by Eq.~(\ref{nfinal}).  Note that this statement is only true when the material has the metric $\bar{g}_{\alpha\beta}$; it is \emph{not} true when the material is distorted into another metric.  In the example at the end of Sec.~V, the material is distorted by some shear strain, and hence it is \emph{not} at the metric $\bar{g}_{\alpha\beta}$.  That is the reason why this example shows a director reorientation driven by the competition of different physical effects.

Note that minimization of the free energy has given Eq.~(\ref{nfinal}) for the director relative to the basis vectors $\vec{t}_\alpha$, as well as Eq.~(\ref{gfinal}) for the metric, which shows the basis vectors relative to each other, but it does not give the individual basis vectors.  This result is physically reasonable, because the material has symmetry under an overall rotation of the director and the basis vectors.  It is impossible to predict the absolute orientation of these vectors in the current space; it is only possible to predict them relative to each other.

If the actual metric $g_{\alpha\beta}$ is slightly different from the reference metric $\bar{g}_{\alpha\beta}$, then the material is strained away from its ground state.  Indeed, the difference between the metric and the reference metric is one definition of the strain tensor, which might be called $\epsilon^\text{pure}$,
\begin{equation}
\epsilon^\text{pure}_{\alpha\beta}=\frac{g_{\alpha\beta}-\bar{g}_{\alpha\beta}}{2}.
\end{equation}
It can be contrasted with the standard Green-Lagrange strain tensor, which might called $\epsilon^\text{apparent}$,
\begin{equation}
\epsilon^\text{apparent}_{\alpha\beta}=\left(\frac{\bm{\lambda}^T \bm{\lambda}-\underline{\underline{I}}}{2}\right)_{\alpha\beta}
=\frac{g_{\alpha\beta}-(g_0)_{\alpha\beta}}{2}.
\end{equation}
Note that $\epsilon^\text{pure}$ compares the current metric with the optimal local metric \emph{at the current temperature and other conditions}.  By comparison, $\epsilon^\text{apparent}$ compares the current metric with the optimal local metric \emph{at the time of crosslinking}.  Hence, $\epsilon^\text{pure}$ is the relevant quantity that shows whether the material is experiencing any local stresses under the current conditions, although $\epsilon^\text{apparent}$ is easier to measure experimentally.

In conclusion, we have developed a formalism that connects the neoclassical theory of liquid crystal elastomers with the recent theoretical approach based on geometric elasticity.  One feature of this formalism is that it keeps track of the director field, and demonstrates that the optimal director is determined by a physical minimization of the free energy rather than by a mathematical transformation of the director at the time of crosslinking.  Assuming that the director is locked into the optimal orientation, the theory calculates the reference metric tensor for strongly crosslinked elastomers, and it shows how this reference metric responds to a change in temperature.  These results should be useful in designing materials for controllable shape changes.

\bigskip
\acknowledgments

We would like to thank H.~Aharoni, E.~Efrati, C.~Modes, E.~Sharon, and M.~Warner for helpful discussions.  This work was supported by National Science Foundation Grant No.~DMR-1409658.

\bibliography{LCE_differential_geometry2}

\end{document}